\documentclass[twoside,fleqn]{article}
\usepackage{espcrc2}
\usepackage{graphicx}
\newcommand{\cD}{{\cal D}}
\newcommand{\wh}{\widehat}
\newcommand{\ep}{\epsilon}

\newcommand{\nn}{\nonumber}

\newcommand{\eqn}[1]{(\ref{#1})}

\newcommand{\gev}{\mbox{\rm GeV}}
\newcommand{\fm}{\mbox{\rm fm}}

\newcommand{\cDt}{\widetilde{\cal D}}

\newcommand{\vacl}{\langle 0|}
\newcommand{\vacr}{|0\rangle }
\newcommand{\prp}{\frac{1}{2}\,(1+\!\!\not{\!v})}
\newcommand{\gluonkond}{\left<aFF\right>}
\newcommand{\expo}{e^{\,gf^{abc}z^\tau \int_0^1 dt A^c_\tau(x+tz)}}

\title{
\vspace{-1.3cm}
{\small\sf
\rightline{HD-THEP-99-26}}
\bigskip
{\Large \bf The field strength correlator from QCD sum rules}}
\author{M. Eidem\"uller\thanks{Invited talk at the Euroconference on Quantum Chromodynamics
        (QCD'99), Montpellier, July 1999}, H.G. Dosch and M. Jamin
\address{
       {\em Institut f\"ur Theoretische Physik, Universit\"at Heidelberg,} \\
       {\em Philosophenweg 16, 69120 Heidelberg, Germany}} 
       }

\begin{document}

\begin{abstract}

The gauge invariant two--point correlator for the gluon field strength
tensor is analysed by means of the QCD sum rule method. To this end, we make
use of a relation of this correlator to a two--point function for a quark--gluon
hybrid in the limit of the quark mass going to infinity. From the sum rules
a relation between the gluon correlation length and the gluon condensate
is obtained.

\end{abstract}

\maketitle

%%%%%%%%%%%%%%%%%%%%%%%%%%%%%%%%%%%%%%%%%%%%%%%%%%%%%%%%%%%%%%%%%%%%%%%%%
% Beginning of the paper
%%%%%%%%%%%%%%%%%%%%%%%%%%%%%%%%%%%%%%%%%%%%%%%%%%%%%%%%%%%%%%%%%%%%%%%%%

\section{Introduction}

An important role in non--perturbative approaches to QCD is played by the
gauge invariant non--local gluon field strength correlator 
\begin{eqnarray}
\label{eq:1}
\lefteqn{\cD_{\mu\nu\rho\sigma}(z)  \equiv  \langle 0|T\{g_s^2 F^a_{\mu\nu}(y)
{\cal P}\expo } \nn\\ &  &\times F^b_{\rho\sigma}(x)\}|0\rangle \,,
\end{eqnarray}
where the field strength $F^a_{\mu\nu}=\partial_\mu A^a_\nu-\partial_\nu
A^a_\mu+gf^{abc}A^b_\mu A^c_\nu$, $z=y-x$ and ${\cal P}$ denotes path
ordering of the exponential. In general, the gauge invariant field strength
correlator could be defined with an arbitrary gauge string connecting the
end points $x$ and $y$, but in this
work we shall restrict ourselves to a straight line.

It is the basic ingredient in the stochastic model of the QCD
vacuum \cite{dos:87,ds:88} and in the description of high
energy hadron-hadron scattering \cite{nr:84,ln:87,kd:90,dfk:94}. In the
spectrum of heavy quark bound states it governs the effect of the gluon
condensate on the level splittings \cite{gro:82,cgo:86,kdb:92,sty:95} and it is
useful for the determination of the spin dependent parts in the heavy quark
potential \cite{sd:88,sim:89}.

The correlator can be related
to a correlator of a colour singlet current composed of a (fictitious)
infinitely heavy octet quark and the gluon field strength tensor. This
fact has already been employed in ref. \cite{ej:98} in order to calculate
the perturbative corrections by means of
Heavy Quark Effective Theory (HQET).
In this work we again use this relation to estimate the correlation length
from QCD sum rules \cite{svz:79} using as ingredients
the value of the gluon condensate and the results for the perturbative
calculation.

\section{The field strength correlator}

Instead of dealing with the non--local string operator which makes a 
calculation rather tedious one can replace the string by local heavy
quark fields and use the methods developed in HQET. To this aim,
we introduce an infinitely heavy quark field in the octett representation,
$h^a(x)$, which is constructed from the field $Q^a(x)$ analogous to HQET by
\begin{equation}
\label{eq:2}
h^a(x)  =  \lim_{m_Q\rightarrow\infty}\,\prp\,e^{im_Qvx}Q^a(x) \,,
\end{equation}
$v$ is the four--velocity of the heavy quark. 
The propagator of the free heavy quark field in coordinate space is given by
\begin{eqnarray}
\label{eq:3}
S(z) &=& \vacl T\{h^a(y)\bar{h}^b(x)\} \vacr \nn\\ &=&  \delta^{ab}
\frac{1}{v^0}\,\theta(z^0)\,\delta\Big({\bf z}-\frac{z^0}{v^0}{\bf v}\Big) \,,
\end{eqnarray}
where  $v^0$ is the zero--component of the velocity. The $\delta$--function
constrains the heavy quark on a straight line.
With the effective HQET action $S_{eff}=\int dx \ \bar{h}\,iv^\mu D_\mu h\,$, the
following equation can be shown analytically \cite{eid:97}:
\begin{eqnarray}
\label{eq:4}
\lefteqn{\vacl T\{h^a(y)\bar{h}^b(x)\,e^{iS_{eff}}\} \vacr}\nn\\  
 & & =  S(z)\,\vacl {\cal P}\expo \vacr \,.
\end{eqnarray}
The physical picture of this result is a heavy quark moving from point
$x$ to $y$ with a four--velocity $v$, acquiring a phase proportional to the
path--ordered exponential. 
The limit of $m_Q\rightarrow\infty$ is necessary in order to constrain the
heavy quark on a straight line and in order to decouple the spin interactions
which are suppressed by a power of $1/m_Q$ and can therefore be neglected. 
By introducing a new correlator $\cDt(z)$, we can now express our correlator \eqn{eq:1}
in terms of heavy quark fields
\begin{eqnarray}
\label{eq:5}
\lefteqn{\cDt_{\mu\nu\rho\sigma}(z)}\nn\\
& \equiv & \vacl T\{g_s^2 F^a_{\mu\nu}(y)h^a(y)F^b_{\rho\sigma}(x)\bar{h}^b(x)
e^{iS_{eff}}\}\vacr \nn \\
& = & S(z) \, D_{\mu\nu\rho\sigma}(z) \,,
\end{eqnarray}
which establishes the relation between the field strength correlator and HQET.

\section{The sum rules}

Our next aim is to evaluate this correlator in the framework of QCD sum rules
\cite{svz:79} and in that way obtain information on the correlation length
of the field strength correlator.
We may view the composite operator $(g_s h^a F^a_{\mu\nu})(x)$
as an interpolating field of colourless quark gluon hybrids and
evaluate $\cDt_{\mu\nu\rho\sigma}(z)$  by introducing these as
intermediate states in the absorption part of $\cDt_{\mu\nu\rho\sigma}(z)$.
The lowest lying state will govern the
long--range behaviour and hence the inverse of its energy is the correlation
length.

For the sum rule analysis it is preferable to work with the correlator in
momentum space. Thus we define
\begin{eqnarray}
\label{eq:6}
\cDt_{\mu\nu\rho\sigma}(w)  &=&  i \int dz \, e^{iqz} \vacl
T\{g_s^2 F^a_{\mu\nu}(y)h^a(y)\nn\\ & &\times F^b_{\rho\sigma}(x)\bar{h}^b(x)\}\vacr \,,
\end{eqnarray}
where $w=vq$ is the residual heavy quark momentum.

For a sum rule analysis the states have to be classified according to
different quantum numbers. The projections onto the two independent
subspaces is done with
\begin{eqnarray}
\label{eq:7}
\cDt^-(w) & \equiv & g^{\mu\rho}v^\nu v^\sigma \, \cDt_{\mu\nu\rho\sigma}(w)\nn\\
\cDt^+(w) & \equiv & (g^{\mu\rho}g^{\nu\sigma}-2\,g^{\mu\rho}v^\nu v^\sigma) \,
\cDt_{\mu\nu\rho\sigma}(w) \,,
\end{eqnarray}
where $\cDt^-$ contains a vector and  $\cDt^+$ an axialvector
intermediate state.

We model the correlators by
a contribution from the lowest lying resonance plus the perturbative
continuum above a threshold $s_0$. Inserting the matrix elements and
performing the heavy quark phase space integrals one obtains
\begin{equation}
\label{eq:8}
\cDt^\mp(w)  =  \frac{\kappa^\mp\,|f^\mp|^2}{w-E^\mp+i\ep} +
\int\limits_{s_0^\mp}^\infty d\lambda \, \frac{\rho^\mp(\lambda)}
{\lambda-w-i\ep} \,,\\
\end{equation}
where $E$ represents the energy
of the glue around the heavy quark, $f^\mp$ are the hadronic matrix
elements and $\kappa^\mp$ are constants. 
The spectral densities are defined by
$\rho^\mp(\lambda)\equiv 1/\pi\,{\rm Im}\,\cDt^\mp(\lambda+i\ep)$ and are
known at the next--to--leading order \cite{ej:98}.

After Fourier transformation to coordinate space the above representation
reads:
\begin{eqnarray}
\label{eq:9}
\cDt(z)  &=&  S(z)\, \biggl\{-\kappa\,|f|^2 e^{-iE|z|}  \nn \\
& & + \int_{s_0}^\infty d\lambda \,
\rho(\lambda) \, e^{-i\lambda |z|} \biggr\} \,.
\end{eqnarray}
Since the heavy quark propagator factorises, we identify the
expression inside the brackets with our original correlator $\cD(z)$.
The long--range behaviour will be dominated by the term containing 
the exponential with the energy $E$. Therefore the correlator decays
exponentially  and the correlation length is given by $1/E$.

Now we turn to the theoretical side
of the sum rules which is obtained by calculating the correlator of
eq.~\eqn{eq:6} in the framework of the operator product expansion
\cite{svz:79,wil:69}.

The perturbative contributions in momentum space have the form
\begin{eqnarray}
\label{eq:10}
\cDt^\mp_{PT}(w) &=& (-w)^3\, a\,\Big[\, p_{10}^\mp+p_{11}^\mp L \nn\\
& & +a\, (p_{20}^\mp+p_{21}^\mp L+p_{22}^\mp L^2) \,\Big] \,,
\end{eqnarray}
where $a\equiv\alpha_s/\pi$, $L=\ln(-2w/\mu)$ and the coefficients
$p_{ij}^\mp$ can be found in ref. \cite{dej:99}.

Essential for the sum rule analysis are the contributions coming from
the condensates. 
In our case
the dimension three condensate $\langle\bar{h}h\rangle$ vanishes since
the quark mass is infinite. The lowest nonvanishing term is the gluon
condensate of dimension four:
\begin{equation}
\label{eq:11}
\cDt^-_{FF}(w) = \frac{1}{2}\,\cDt^+_{FF}(w)=  -\,\frac{\pi^2}{w}\gluonkond\,.
\end{equation}
The next condensate contribution would be of dimension six, but we shall
neglect all higher condensate contributions in this work and restrict
ourselves to the gluon condensate.

The correlators satisfy homogeneous
renormalisation group equations. Thus we can improve the perturbative
expressions by resumming the logarithmic contributions. 
Calculation of the first coefficients $\gamma_1^\mp$ for the
anomalous dimensions from eq. \eqn{eq:10} leads to 
\begin{equation}
\label{eq:12}
\gamma_1^- \; = \; 0 \,, \qquad
\gamma_1^+ \; = \; 3 \,.
\end{equation}
The correlator $\cDt^-(w)$ which corresponds to the
vector intermediate state does not depend on the renormalisation scale $\mu$
at this order.

In order to suppress contributions in the dispersion integral coming from
higher exited states and from higher dimensional condensates, it is convenient
to apply a Borel transformation $\wh{B}_T$ with $T$ being the Borel variable.
After renormalisation group improvement and Borel transformation
all the ingredients needed for a sum rule analysis are known. Explicit
formulas for the expressions can be found in ref. \cite{dej:99}.
Now we turn to the numerical analysis.

\section{Numerical analysis}

Let us denote by $\chi(T,s_0)$ the Borel transformed expression for the
continuum part $\chi(T,s_0)\equiv\wh{B}_T \int_{s_0}^\infty d\lambda\ \rho(\lambda)
/(\lambda-\omega-i\ep)\,$.
After equating the phenomenological and the theoretical part we end up with
the sum rule
\begin{eqnarray}
\label{eq:13}
\lefteqn{-\kappa^\mp |f^\mp|^2 e^{-E^\mp/T}}\nn\\  
 & & = \wh\cD_{FF}^\mp + \wh\cD_{PT}^\mp(T)-\chi^\mp(T,s_0) \,.
\end{eqnarray}
By taking the logarithmic derivative we get an equation for the
energy $E$.

The analysis shows that the different sign of the perturbative and non--perturbative term
in the $1^-$ state leads to a stabilisation for the energy sum rule, whereas 
the equal sign in the $1^+$ state destabilises.

As our input parameters for the case of three light quark flavours we use
$<aFF>=0.024 \pm 0.012 \ \gev^4$, $\Lambda_{3fl} =0.323
 \ \gev$ and $\mu=2\ \gev$.
To estimate the errors we have varied the scale $\mu$, 
the continuum threshold $s_0$ and the gluon condensate. In Fig. 1 we have 
displayed the energy $E^-$
as a function of the Borel parameter $T$ for different values of $\mu$
and $s_0$. A good balance between stability for the energy and sensitivity
for the resonance is found around $E^-=1.5$ GeV.

% Figure
\begin{figure}[thb]
\includegraphics[angle=-90,width=7.5cm]{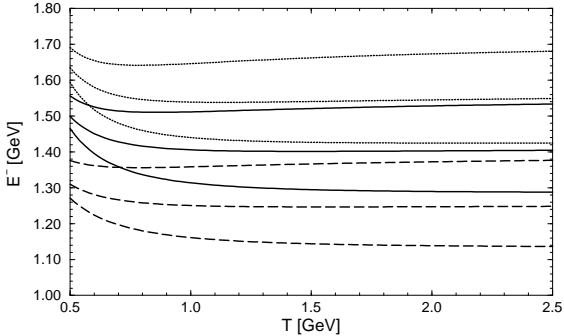}
\caption[]{The energy as a function of the Borel-parameter T.
Dashed curves $\mu\!=\!1\,$GeV: lowest $s_0\!=\!1.3\,$GeV, middle $s_0\!=\!1.5\,$GeV, 
upper $s_0\!=\!1.7\,$GeV. 
Solid curves $\mu\!=\!2\,$GeV: lowest $s_0\!=\!1.5\,$GeV, 
middle $s_0\!=\!1.7\,$GeV, 
upper $s_0\!=\!1.9\,$GeV.
Dotted curves $\mu\!=\!4\,$GeV: lowest $s_0\!=\!1.7\,$GeV, middle $s_0\!=\!1.9\,$GeV, 
upper $s_0\!=\!2.1\,$GeV.
\vspace*{-6mm}
\label{fig:1}}
\end{figure}
Including the errors we
get for the energy and correlation length
\begin{equation}
\label{eq:14}
E^-_{3fl} = 1.5 \pm 0.4 \ \gev, \ a^-_{3fl} = 0.13^{+0.05}_{-0.02} \ \fm.
\end{equation}
In a world without light quarks, i.e. $n_f = 0$, the main influence is the expected 
change of the gluon condensate which might increase by a factor two to 
three \cite{nsvz:81}. 
If we perform an analysis as above, we get for $\Lambda_{0fl} =0.250 \ \gev$, $<aFF>=0.0
48 \pm 0.024 \ \gev^4$  and $s_0=2.3\ \gev$ an energy and correlation length of 
\begin{equation}
\label{eq:15}
E^-_{0fl} = 1.9 \pm 0.5 \ \gev \ , \ a^-_{0fl} = 0.11^{+0.04}_{-0.02} \ \fm.
\end{equation} 
For $E^+$, the energy of the axial vector, we obtain no stable sum rule. The expressions
for $E^-$ and $E^+$ are equal in lowest order perturbation theory,
higher order perturbative contributions and the gluon condensate lead to
a splitting in such a way that for the same values of $s_0$ and $T$ the expression for 
$E^-$ is higher than that for $E^+$.

\section{Summary and conclusions}

The analysis of the gauge invariant gluon field strength correlator
by QCD sum rules allows to establish a relation between the gluon 
condensate and the correlation length. In order to apply the sum rule
technique which consists in the comparison of a phenomenological ansatz
with a theoretical expression obtained from operator product expansion
we interpret the gluon correlator as the correlator of two colour 
neutral hybrid states composed of a (fictitious) heavy quark transforming
under the adjoint representation and a gluon field. The former serves as
the source for the gauge string in the correlator.

The value of the lowest intermediate $1^-$ state (the inverse correlation
length of the correlator) with three flavours can be determined to 
$E^-_{3fl}=1/a^-_{3fl}\approx 1.5 \pm 0.4 
\ \gev$ and with zero flavours to $E^-_{0fl}=1/a^-_{0fl}\approx 1.9 \pm 0.5
\ \gev$. The main sources of uncertainty are the choice of the continuum threshold $s_0$
 and the value of the gluon condensate. 

Though we find no stable sum rule for the axial vector state we have from the difference
 of the expressions for the $1^-$ and $1^+$ state strong evidence for the 
counterintuitive result that the $1^+$ state is lighter than the vector state. 

The field strength correlator has been calculated on the lattice
using the cooling technique \cite{gmp:97,egm:97} and field insertions
into a Wilson loop \cite{bbv:98}. For a discussion see ref. \cite{dej:99}.
Recently in an analysis of heavy quarkonium in the framework of NRQCD 
\cite{bpsv:99} a splitting between the vector and the axialvector
part has been observed in the same direction as proposed by the sum rules.

The sum rule analysis shows that the state investigated here namely a gluon confined by
 an octet source has a much higher energy than a corresponding state in HQET. A similar
 analysis of a light quark bound by a source in the fundamental representation 
\cite{bbbd:92} yielded an energy which is by a factor 2 to 4 smaller. This is to be expected from
general grounds \cite{nsvz:81}  since the case treated here is nearer to a glueball than
to a heavy meson.

\bigskip \noindent
{\bf Acknowledgments}
The authors would like to thank S. Narison for the invitation to
this very pleasant and interesting conference.

\end{document}